\documentstyle[apb_cite,epsf,rotate]{mn}

\title[Investigating the structure of the accretion disk in WZ~Sge.] 
{Investigating the structure of the accretion disk in WZ~Sge from 
multi-wave-band, time-resolved spectroscopic observations: {\bf Paper II}}

\author[Mason {\rm et al.}] {Elena Mason,$^{1,2}$  Warren Skidmore,$^{3,1}$ 
Steve 
B. Howell,$^{1,2}$ David R. Ciardi,$^{4}$ \cr Stuart Littlefair,$^{5}$ V. S. 
Dhillon,$^{5}$ \\
$^{1}$Department of Physics and Astronomy, University of Wyoming, P.O. Box 
3905, University Station, Laramie, WY 82071, USA.\\
$^{2}$Astrophysics Group, Planetary Science Institute, 620 N. 6$^{th}$Ave., 
Tucson, AZ 85705, USA.\\
$^{3}$School of Physics and Astronomy, University of St. Andrews, North 
Haugh, St. Andrews, Fife, KY16 9SS, UK.\\
$^{4}$Department of Astronomy, 211 Bryant Space Science Center, P.O. Box 
112055, University of Florida, Gainesville, FL 32611, USA.\\
$^{5}$Department of Physics and Astronomy, University of Sheffield, Sheffield, 
S3 7RH, UK.}

\begin{document}

\maketitle

\begin{abstract}
 
We present our second paper describing multi-wave-band, time-resolved 
spectroscopy of WZ~Sge. We analyze the evolution of both optical and IR emission 
lines throughout the orbital period and find evidence, in the Balmer lines, for 
an optically thin accretion disk and an optically thick hot-spot. Optical and IR 
emission lines are used to compute radial velocity curves. Fits to our radial 
velocity measurements give an internally inconsistent set of values for K1, 
$\gamma$, and the phase of red-to-blue crossing. We present a 
probable 
explanation for these discrepancies and provide evidence for similar behaviour 
in other short orbital period dwarf-novae. Selected optical and IR spectra are 
measured to determine the accretion disk radii. Values for the disk radii are 
found to be strongly dependent on the assumed WD mass and binary orbital 
inclination. However, the separation of the peaks in the optical emission line 
(i.e. an indication of the outer disk radius) has been  found to be constant 
during all phases of the super-cycle period over the last 40 years. 

\end{abstract}


\section{Introduction}

Dwarf-novae, among the cataclysmic variables (CVs), are semi-detached binary 
star systems where a white-dwarf (primary star) accretes gas from a cool 
companion (secondary star). The gas from the secondary star accretes through the 
inner Lagrangian point (L1), into the primary Roche lobe and slowly spirals in 
toward the white-dwarf. The in-flowing gas forms an accretion disk which orbits 
around the white-dwarf before actually being accreted onto the primary star. 
During the time the gas is within the accretion disk, current theory tells us 
that half of the gravitational energy of the gas is radiated outward while that 
which remains heats the accretion disk. Indeed most of the 
visible and IR light 
output by a typical dwarf-novae is from the accretion disk (\ncite{Ciardi98}, 
\ncite{warner95}). Emission lines seen in the spectra of a dwarf-nova arise from 
the vicinity of the accretion disk and can be double peaked due to the Doppler 
motions of the disk within the binary \cite{Horne86b}. The velocity of the 
material within the disk, together with the emitted flux distribution over the 
disk, determines the line profile shape. Consequently, spectral line analysis is 
a fundamental method employed to understand the physical state of the accretion 
disk and its spatial flux distribution. The most common emission lines used in 
such analyses are those due to hydrogen and helium. 

\begin{table*}
\caption{Log of the observations of WZ~Sge used in this analysis. Velocity 
dispersion is stated for the line used; it is not the mean velocity dispersion 
of the complete data set. ($\dagger$ average value for all three lines). }
 
\begin{center}
\begin{tabular}{cclcccccc}
Date of & Wavelength & Emission & Dispersion & App. dispersion & Exposure time 
& Run duration & Orbits & Number of \\
observations & range (\AA) & Lines & (\AA/pixel) &  (km sec$^{-1}$ ) & 
(sec) & (hours) & covered & spectra \\
\hline
27/07/96 & 6375 - 6780 & H$\alpha$ & 0.40 & 18.1 & 40 & 5.51 & 4.1 & 237 \\
27/07/96 & 4590 - 4995 & H$\beta$ & 0.40 & 24.6 & 40 & 5.03 & 3.7 & 239 \\
28/07/96 & 6375 - 6780 & H$\alpha$ & 0.40 & 18.1 & 40 & 4.07 & 3.0 & 139 \\
28/07/96 & 4590 - 4995 & H$\beta$ & 0.40 & 24.6 & 40 & 4.10 & 3.0 & 137 \\
28/05/97 & 18000 - 24600 & Br$\gamma$, Br$\delta$, He~I & 13.2 
& 193$\dagger$  & 480 & 2.70 & 2.0 & 18 \\
08/08/98 & 10300 - 13400 & Pa$\beta$ & 6.22 & 145.6 & 240 & 6.73 & 3.5 & 30 \\
\end{tabular}
\end{center}
\protect
\label{tab:observations} 
\end{table*}

\begin{figure*}
\begin{minipage}{18cm}
\vspace{10.5cm}
\includegraphics{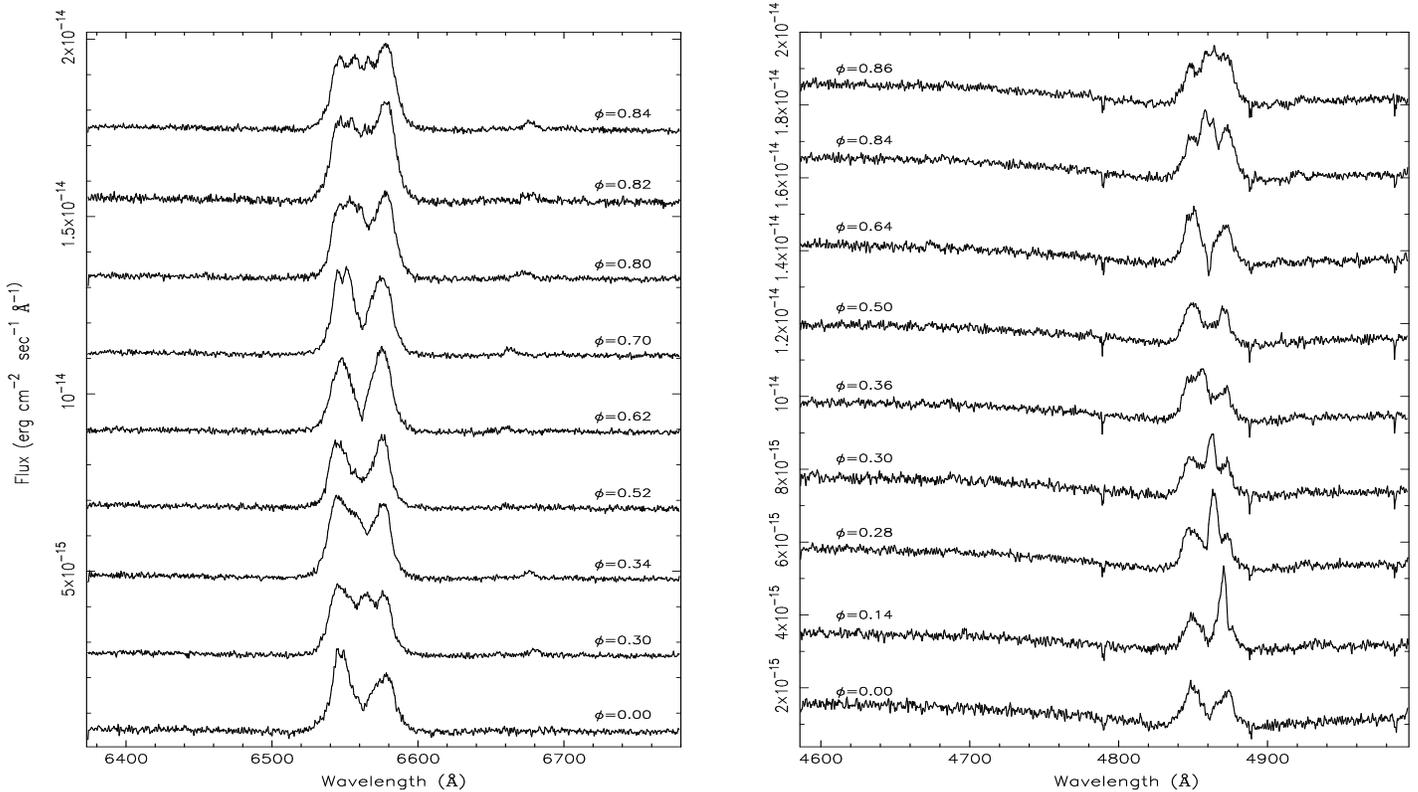}
\caption{ Sample H$\alpha$ and H$\beta$ phase binned spectra (50 bins per 
orbit). Each spectrum is offset from the one below by 
$2.1\times10^{-15}$~erg~cm$^{-2}$~s$^{-1}$~\AA$^{-1}$ and each bin is labeled 
with mid-orbital bin phase based on the optical photometric eclipse ephemeris 
listed in PAPER~I. Note that the H$\alpha$ spectra show a weak 
He~I emission line ($\lambda$ 6678\AA), which follows the H$\alpha$ S-wave 
motion (see trailed-spectrum in figure 1 of PAPER~I). The narrow ``absorption 
features'' seen in the H$\beta$ spectra are due 
to bad pixels. {\it Note:} the left and right panels have the same flux units. }
\label{fig:specplot}
\end{minipage}
\end{figure*}

\begin{figure*}
\begin{minipage}{18cm}
\vspace{10.7cm}
\includegraphics{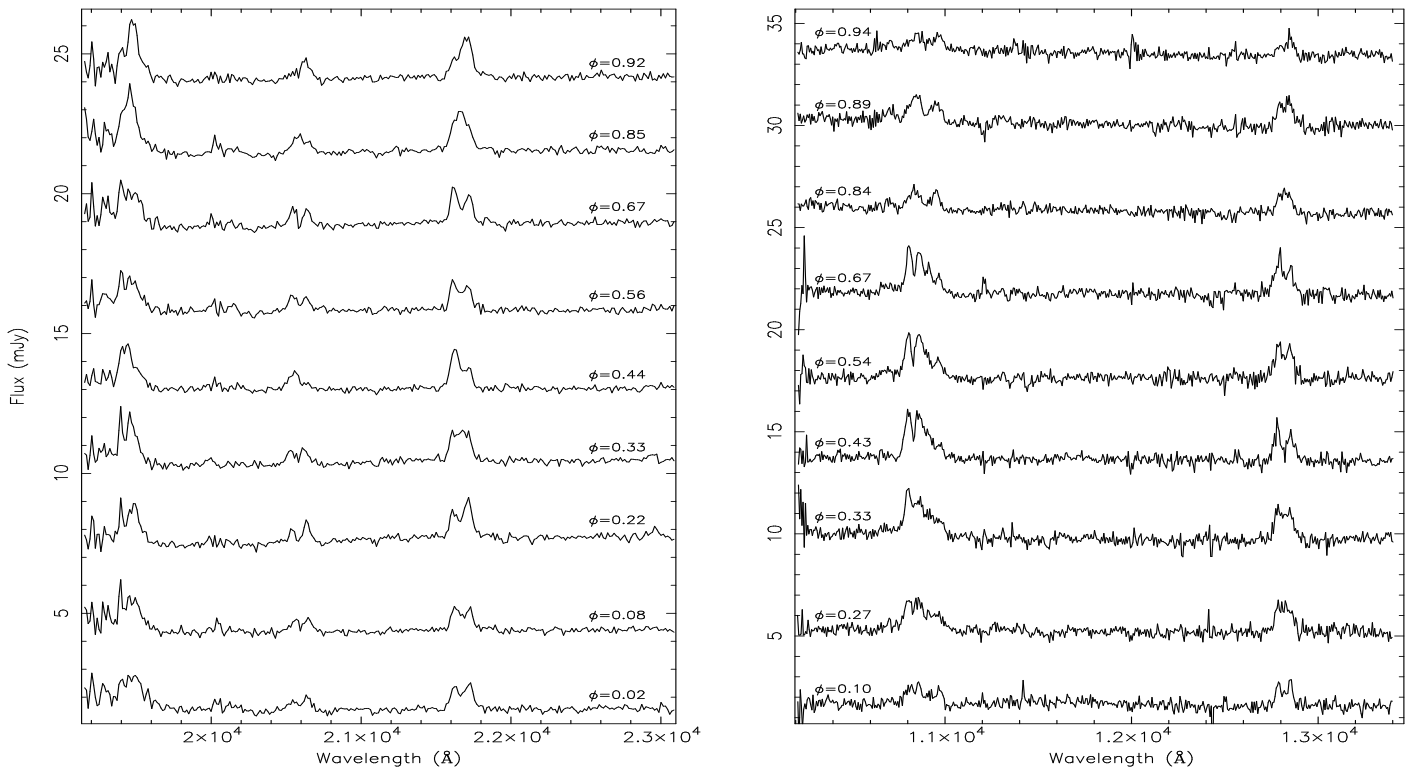}
\caption{ IR spectra in the K and J bands. Spectra are labeled with the 
phase at mid-exposure time as set by the photometric eclipse ephemeris listed in 
PAPER~I. Flux is in mJy and the offset between each spectra 
is 2.8 and 4.0 mJy for K and J spectra respectively. {\it Note:} the left and 
right panels have the same flux units. }
\label{fig:irsample}
\end{minipage}
\end{figure*}

\begin{figure*}
\begin{minipage}{18cm}
\vspace{11cm}
\includegraphics{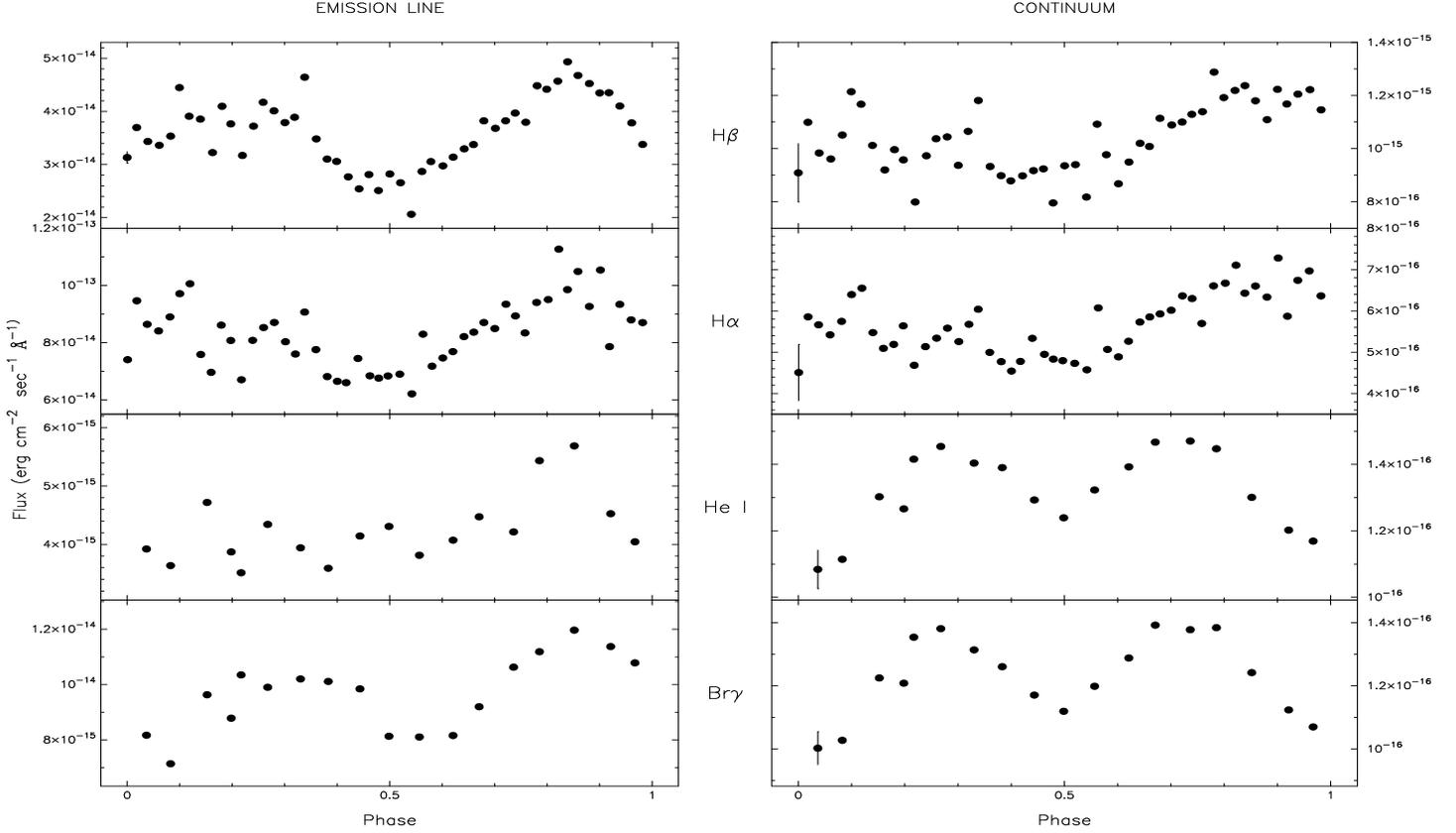}      
\caption{Left: the line flux for each emission line vs the orbital phase; Right: 
the continuum flux underlying the emission lines vs orbital phase. Fluxes 
measured at phase  0.3, 0.5, and 0.84 are in table~\ref{tab:hspercent}. Both 
emission line and continuum fluxes were measured using the {\it e-e} command in 
IRAF task {\it splot}; fitting was not attempted. Average uncertainties on 
measurements are plot on the first data point in each panel; error-bars on 
H$\alpha$, Br$\gamma$, and He~I 2.06$\mu$m emission lines are smaller than the 
data point size.  
{\it Note:} the left and right panels have the same flux units. }
\label{fig:fluxes}
\end{minipage}
\end{figure*}

\begin{figure}
\vspace{9.0cm}           
\includegraphics{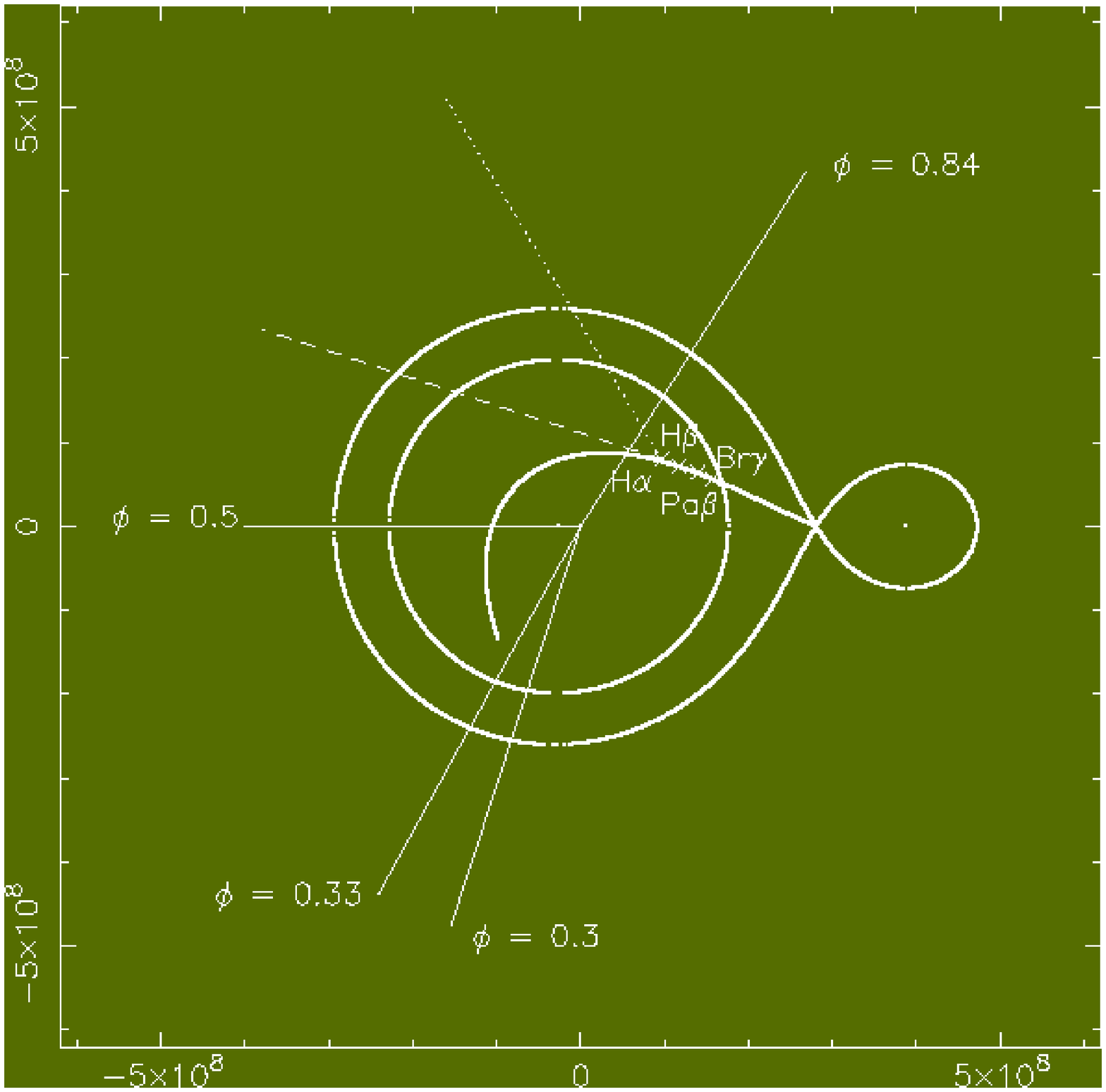}
\caption{The Roche lobe geometry of WZ~Sge corresponding to a white-dwarf mass 
of 0.82M$_\odot$ (PAPER~I), and the secondary star mass (0.058M$_\odot$) 
computed by \protect\scite{Smak}. Units on both the x and y axis are m. 
Straight full lines mark the line of sight at the phases considered in the text. 
Dashed and dotted lines show the directions of stream and Keplerian velocity 
respectively for the illustrative case of H$\alpha$ hot-spot position. Crosses 
along the stream trajectory qualitatively mark the hot-spot position for each 
emission line as predicted by our results. From the left to the right: 
H$\alpha$, H$\beta$, Pa$\beta$, and Br$\gamma$ in order of increasing excitation 
energy and gas temperature. }
\label{fig:geo}
\end{figure}

In this paper, we continue the analysis of the accretion disk in WZ~Sge which 
was started in \ncite{pap1} (hereafter PAPER~I). PAPER~I presented Doppler 
tomography of optical and IR disk emission lines as well as ratioed Doppler 
maps. General conclusions reached there indicated different optical depth at the 
hot-spot and the accretion disk, and an accretion disk structure dissimilar to 
those predicted by models. In this paper (PAPER~II), we perform spectral 
analysis by direct measurements of the emission line profiles. We organize the 
present paper in three main sections. Section 2 describes the orbital phase 
dependence of H$\alpha$, H$\beta$, and IR emission lines. Section 3 presents 
newly determined radial velocity curves using optical and IR emission lines. We 
show that the amplitude and phase zero of the radial velocity curves are 
systematically biased. We show that this is due to different gas opacities in 
the hot-spot and the accretion disk, the hot-spot having a larger opacity than 
the accretion disk. We further develop the idea of increasing hot-spot 
contamination in systems where the difference between the optical depth in the 
hot-spot and the accretion disk is large, these systems being short orbital 
period, low mass transfer rate dwarf-novae. In section 4 we present our 
determinations for the accretion disk radii and compare our values with 
previously reported determinations. 

Complete details concerning the data set used here (see table 
\ref{tab:observations}), are in PAPER~I.
All data are phased according to the 
photometric eclipse ephemeris stated in table~7, PAPER~I. Optical spectra of 
WZ~Sge were grouped into 50 phase bins per orbit in order to improve the 
signal-to-noise ratio of the data. 

\section{Emission lines}
\label{sec:lp}

We analyze the emission lines evolution by inspecting their profile, measuring 
the fluxes and computing the Balmer decrement (last analysis on optical spectra 
only). Figure~\ref{fig:geo} summarizes the results from each analysis. 

\subsection{The line profiles}

Figures \ref{fig:specplot} and \ref{fig:irsample} present a sample of 
optical and IR spectra respectively to show the complex emission line profiles 
and their dramatic evolution throughout an orbital period. Both optical and IR 
emission line profiles evolve similarly and owe most of their changes to the 
variable strength of the hot-spot around the orbit. Phases 0.3, 0.5, 0.84 and 
0.63 are particularly suitable to point out the hot-spot evolution over the 
orbit. 

Inspection shows that phase 0.3 is when the hot-spot, in both H$\alpha$ and 
H$\beta$ emission lines, is at the center of the accretion disk emission, i.e. 
has zero radial velocity with respect to an observer co-moving with the 
white-dwarf. From the system geometry determined in PAPER~I we expect the  
hot-spot in the middle of the accretion disk 
emission line at phase $\sim$0.3 
{\it only} if the gas at the stream-disk impact region has just the stream 
velocity (see figure~\ref{fig:geo}). This condition implies that the accretion 
disk density is low enough such that the gas at the stream-disk impact 
region keeps its stream velocity with little effect from the Keplerian motion of 
the underlying accretion disk material. In the IR spectra the phase of the 
hot-spot zero radial velocity occurs slightly later, i.e. at phase 0.33. Phase 
binning of the optical spectra with the same time resolution of the IR spectra 
has showed that such a delay is significant and implies different velocity of 
the hot-spot gas in different wave-bands. We explain such a phase delay with the 
hot-spot emission from different lines spreading along the stream trajectory 
(figure~\ref{fig:geo}).

The hot-spot appears again at the center of the emission line at phase 0.84, 
both in the optical and IR spectra. The expected phase for the optical spectra 
is 0.80 in the assumption of symmetric and isotropic hot-spot emission. Further 
evidence for asymmetric and anisotropic hot-spot emission is in the hot-spot 
profile which appears double peaked at phase 0.84. This feature is real as it is 
common to both H$\alpha$ and H$\beta$ {\it binned} spectra and is visible for a 
phase interval of $\leq$0.5 (see also the trailed spectrograms in figure 1 of 
PAPER~I). A detailed study of the double structure is beyond the 
scope of this paper. The lower spectral resolution in the J and K bands does not 
resolve such a feature if present (the hot-spot peak separation is $\simeq$9\AA 
\ in the Balmer lines). 

Phase $\sim$0.53 is when we expect the maximum hot-spot blue-shift and enhanced 
blue peak emission in the accretion disk lines. Observations do not confirm this 
expectation and show red and blue shifted peaks of roughly equal strength  
between phase 0.5 and 0.56, both in the optical and IR spectra. We explain this  
observation with little or no contribution from the hot-spot to the overall 
emission line flux. This is also evidence of the anisotropic hot-spot emission. 

Asymmetries in the accretion disk are evident in the H$\alpha$ emission line 
which displays a stronger red-shifted peak at phase 0.63. Doppler maps in 
PAPER~I show asymmetric accretion disks also in the 
H$\beta$, Br$\gamma$, He~I, 
and Br$\delta$ lines. The accretion disk emission line profile evolves from a 
shallow U-shaped profile to a V-shaped profile, between phase 0.5 and 0.63. This 
is particularly evident in the optical spectra; among IR observations, only the 
J band spectra show similar evolution. \scite{Horne86b} predict U-shaped 
emission line profiles for optically thin accretion disks, and V-shaped profiles 
for the optically thick case. This might suggest that the changing profile of 
the emission lines is evidence that some areas of the disk have optically thin 
line emission and some have optically thick. 

\subsection{Line flux}
\label{sec:lflx}

\begin{table}
\begin{center}
\begin{tabular}{cccc}
Line & Max Flux  & Min Flux  & Flux at $\phi\sim0.3$ \\
\hline
H$\alpha$ & 11.00$\times$10$^{-14}$ & 6.75$\times$10$^{-14}$ & 
8.80$\times$10$^{-14}$ \\
H$\beta$ & 4.95$\times$10$^{-14}$ & 2.65$\times$10$^{-14}$ & 
4.00$\times$10$^{-14}$ \\
Br$\gamma$ & 1.19$\times$10$^{-14}$ & 1.01$\times$10$^{-14}$ & 
8.32$\times$10$^{-15}$ \\
He~I 2.06$\mu$m & 5.67$\times$10$^{-15}$ & 3.54$\times$10$^{-15}$ & - \\
\end{tabular}
\caption{The maximum and minimum fluxes, and the flux near phase 0.3, as 
measured in each emission line. Flux units are erg cm$^2$ sec$^{-1}$ 
\AA$^{-1}$. The minimum fluxes are observed about phase 0.5 and are believed to 
correspond to the phase when there is no contribution from the hot-spot. The 
maximum flux is observed at phase 0.84 when the hot-spot is viewed from the 
outside. Phase 0.3 is the phase when the hot-spot is viewed from the inside.}
\label{tab:hspercent}
\end{center}
\end{table}

The emission line flux was measured by manually marking the position of line 
wings with the cursor and summing up the flux from each pixel in between; the 
continuum flux corresponding to a linear interpolation of the same two cursor 
position was automatically subtracted during the line flux measurements.
Each of the emission lines measured (with the exception of the Pa$\beta$ line) 
show quite strong modulations apparently correlated with the line profile 
evolution (figure~\ref{fig:fluxes}). The modulations display maxima at phase 
0.84 and 0.3, and a minimum at phase 0.5. Thus, the hot-spot shows stronger 
emission when viewed from the outside (phase 0.84) and from the inside (0.3), 
while it does not contribute significantly to the emission line flux when 
viewed from upstream (phase 0.5). Assuming that the flux at phase 0.5 is due 
only to the accretion disk (see figure~\ref{fig:specplot} and figure 
\ref{fig:irsample}, and previous subsection) we can compute the relative 
contribution by the hot-spot to the total emission line flux. At phase 0.84 the 
hot-spot emits 50$\%$ of the total line flux in H$\beta$ and $\sim30\%$ in 
H$\alpha$ and the IR lines. When, viewed form the inside the hot-spot relative 
contribution decreases to 38$\%$ in H$\beta$ and to $\sim20\%$ in H$\alpha$ and 
Br$\gamma$. The He~I 2.06$\mu$m line does not display an evident hump at phase 
0.3, thus the relative hot-spot contribution was not computed there. 
Assuming a spherical hot-spot with radius $S=3.1\times10^{8}$cm \cite{Smak}, 
cylindrical accretion disk with thickness $2H\sim3.8\times10^{8}$cm 
\footnote{The value was derived computing the scale height H as given in 
\cite{will88} assuming a uniform temperature T=6000K in the accretion disk and a 
white dwarf mass of M$_{WD}=0.82$M$_\odot$ (PAPER~I).}, and inclination $i=75^o$ 
\cite{Smak}, the decrease of the hot-spot flux from phase 0.84 to phase 0.30 due 
to occultation by an optically thick disk is 70\%. We observe a decrease of only 
30\% in H$\beta$ and $\sim$50\%  in H$\alpha$ and Br$\gamma$. Thus either the 
accretion disk gas is at least partially transparent to the hot-spot emission 
line flux, or the assumed geometry inappropriate. 

The continuum light curve is expected to be double-humped as already observed by 
\scite{Ciardi98} in the IR, and by \scite{patt96} and 
\scite{skidthesis} in the 
optical. Right panel of figure~\ref{fig:fluxes} shows that only our K-band 
continuum measurements match with previous observations. The optical continuum 
shows large uncertainties on each data point and poor modulation throughout the 
orbit as the method used to measure line and continuum flux does not take into 
account the underlying white-dwarf absorption. The IR spectra, which do not show 
any evidence of white-dwarf features have the expected double-humped structure 
and smaller uncertainties in the continuum light curves. Fluxes from the 
Pa$\beta$ emission line and its underlying continuum have not been reported in 
figure~\ref{fig:fluxes} and table~\ref{tab:hspercent} because the data show 
variations throughout the run of observation. The run covered about four cycles 
and variations between cycles do not appear to be correlated. We cannot say 
whether this is due to intrinsic variability of the emitting source or to 
external causes. 

\subsection{The Balmer decrement}

\begin{figure}
\vspace{10cm}           
\includegraphics{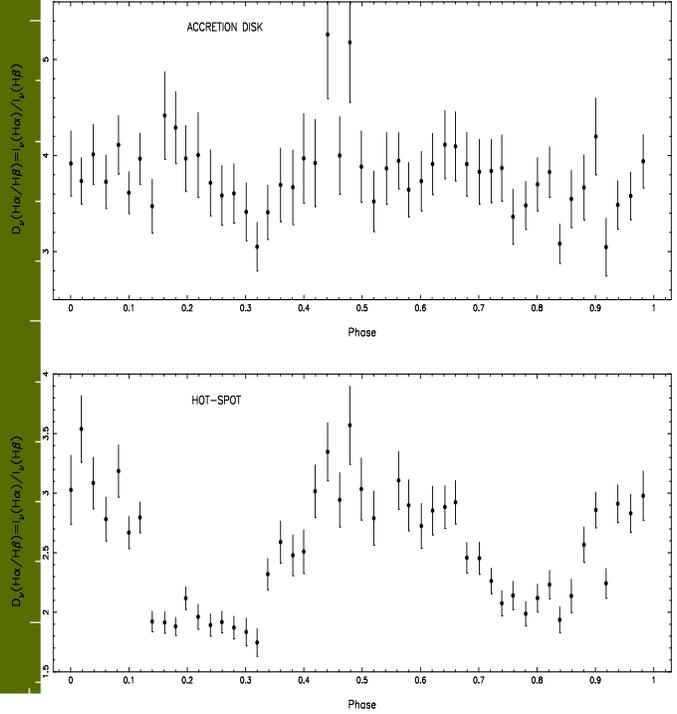}
\caption{The Balmer decrement, ${\it D}_\nu(H\alpha/H\beta)$, versus orbital 
phase both for the accretion disk (top) and hot-spot (bottom). }
\label{fig:iratio}
\end{figure}

We measured the Balmer decrement for both the accretion disk and the hot-spot 
region to check for different opacities. The Balmer decrement is defined by 
the ratio of the line intensity  in frequency units ${\it D}_\nu(H\alpha/H\beta) 
\equiv I_\nu (H\alpha)/I_\nu (H\beta)$. The Balmer decrement, as measured from 
spectra, relies on line fluxes, because the line intensity is broadened by 
effects due to both the intrinsic properties of the emitting gas and the 
instrumentation used during the observation. The comparison of measured 
flux-ratios to tabulated intensity-ratios assumes that the emission lines of 
interest have identical profiles, each broadened in a similar way. We observe 
Balmer lines which are far from being identical in shape, however, table 6 of 
PAPER~I show that time averaged disk profiles have {\it identical} velocity 
broadening. De-blending of the components is not possible without a complex 
model of hot-spot shape and accretion disk structure, thus we measured the peak 
intensity for each component, on continuum-subtracted spectra. We measured the 
disk emission at each phase using only the red or blue line peak, whichever was 
unaffected by the hot-spot emission at the given phase. The hot-spot line 
intensity has always some underlying disk emission blending, except near phase 
$\sim0.3$ and $\sim0.84$, when the hot-spot emission is at the center of the 
emission line and the relative contribution from the accretion disk is 
negligible. The top panel of figure~\ref{fig:iratio} shows that the Balmer 
decrement within the accretion disk is approximately constant throughout the 
orbit with an average value of ${\it D}_\nu (H\alpha/H\beta)=3.82$, which is 
larger than the predicted value of ${\it D}_\nu (H\alpha/H\beta)\sim 1$ for an 
optically thick accretion disk (\ncite{bob}, \ncite{williams83}). 
$I_\nu(H\alpha)$ and $I_\nu(H\beta)$ each show phase dependent modulations, 
implying a non-uniform accretion disk emission, but their ratio remains 
constant. However, despite the uncertainties, the Balmer decrement in the 
accretion disk shows a clear trend around phase 0.3 toward higher opacities. The 
bottom panel of figure~\ref{fig:iratio} shows the hot-spot Balmer decrement vs 
orbital period. The Balmer decrement here is strongly phase dependent and shows 
two deep minima at phase intervals 0.2-0.3 and 0.8-0.84, i.e. when observing 
perpendicular to the stream trajectory at the hot-spot position and measuring 
the hot-spot line intensities without any underlying accretion disk emission. 
The hot-spot Balmer decrement is ${\it D}_\nu (H\alpha/H\beta)\sim$2. The value 
of ${\it D}_\nu (H\alpha/H\beta)$ outside phases 0.2-0.3 and 0.8-0.84 is $\sim$ 
3 and approaches the average accretion disk Balmer decrement  when the hot-spot 
emission is weak. In the hot-spot,  $I_\nu(H\beta)$ shows a stronger modulation 
around the orbit than $I_\nu(H\alpha)$ which is largely biased by the underlying 
accretion disk emission.

\begin{figure}
\vspace{9cm}           
\includegraphics{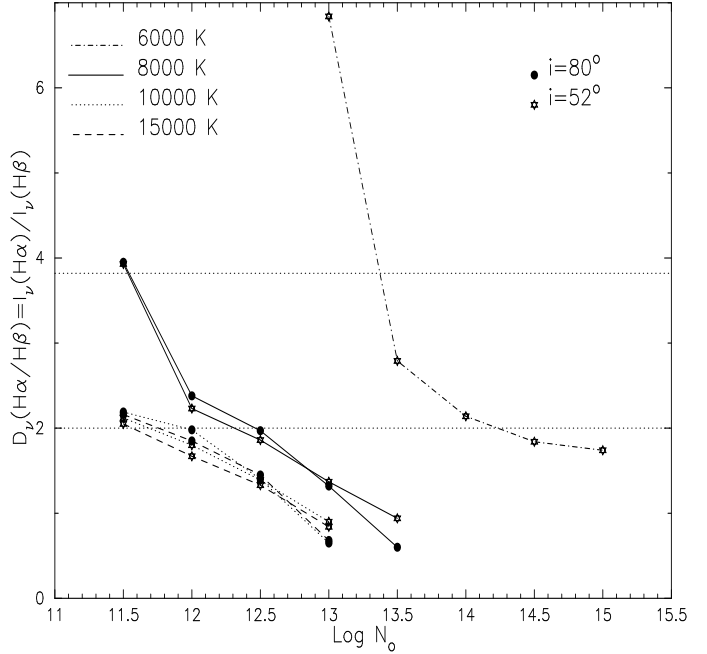}
\caption{Balmer decrement, ${\it D}_\nu(H\alpha/H\beta)$, as predicted by 
\protect\scite{williams} for optically thin disk emission lines. Each line 
represents the Balmer decrement as function of the mid-plane accretion disk 
density, $N_o$ cm$^{-3}$, for a given temperature and orbital inclination. The 
models for T=8000, 10000, and 15000K, all approach ${\it 
D}_\nu(H\alpha/H\beta)\sim1$ at high density ($\log N_o\sim 13.-13.5$), as 
expected for an optically thick gas. The horizontal lines mark the Balmer 
decrement measured for the accretion disk (${\it D}_\nu(H\alpha/H\beta)=3.82$) 
and the hot-spot (${\it D}_\nu(H\alpha/H\beta)=2$) in WZ~Sge. }
\label{fig:glenmodel}
\end{figure}

The Balmer decrement provides information about the gas temperature and density 
where the emission line forms; it gives a unique gas density for each assumed 
temperature. However, care must be taken because different temperatures may give 
the same value of the Balmer decrement depending on the assumption of optically 
thick or optically thin emission lines. In the interpretation of our results we 
mainly use models by \scite{williams} who modeled emission lines in optically 
thin accretion disks. He computed H$\beta$ strengths and Balmer decrements for a 
grid of temperatures, orbital inclinations, and 
mid-plane nucleon densities. In 
figure~\ref{fig:glenmodel} we plot the results of \scite{williams} together with 
the Balmer decrement we measure at the hot-spot (lower horizontal line) and the 
accretion disk (upper horizontal line) in WZ~Sge.

The value of ${\it D}_\nu (H\alpha/H\beta)$=2 for the hot-spot region 
corresponds to a black-body optically thick emission with a temperature of 
T$<$5000K. This decrement apparently matches a number of observed Balmer 
decrements in CVs \cite{williams83}. However, because we observe the He~I 
emission line at 2.06$\mu$m, parts of the hot-spot region must be at least 
15000K (see section~\ref{sec:radvel}). An optically thin gas with a temperature 
of 15000K and a Balmer decrement of 2 corresponds to a gas density of $\log 
N_o=11.5$ in Williams' model (see figure~\ref{fig:glenmodel}), which is lower 
than the derived gas density within the accretion disk (see below). The gas 
density in the stream is expected to be 3 to 4 times larger than the gas density 
within the accretion disk \cite{lu76}. We thus conclude that the hot-spot is at 
least partially optically thick and characterized by a steep temperature 
gradient. The Balmer emission lines form down-stream from the initial impact 
region from a relatively low temperature (T$\sim$5000K) optically thick gas, and 
the He~I 2.06$\mu$m emission line forms in the outer stream-impact region where 
the temperature is higher (see section~\ref{sec:radvel}).

The Balmer decrement ${\it D}_\nu (H\alpha/H\beta)$ = 3.82 of the gas within the 
accretion disk would correspond to a black-body of temperature T$\sim$3400K. 
However, we expect the temperature to be higher as in an optically thick 
$\alpha$-disk the temperature is never expected to fall below 6000K over a wide 
range of mass accretion rate \cite{bob}. Moreover, Ratioed Doppler tomograms and 
radial disk profile of WZ~Sge (see PAPER~I) suggest a partially optically thin 
accretion disk. We conclude that the accretion disk is optically thin in the 
Balmer emission lines. According to \scite{williams} a Balmer decrement of  
${\it D}_\nu (H\alpha/H\beta)$ = 3.82 
corresponds to an optically thin gas of 
temperature 6000K and density $\log N_o \sim$ 13.5 (see figure 
\ref{fig:glenmodel}). A roughly similar value of $\log N_e\simeq 13$ for gas at 
a temperature of T=5000K may be derived from \scite{darke} models. 
\scite{darke}, using the escape probability approach, determined Balmer 
decrements and hydrogen lines strength for an infinite slab of gas at various 
opacities, temperatures, and densities. We can use these estimates to determine 
the optically thin accretion disk mass assuming: {\it i)} a uniform gas density 
of $\log N_o=13.2$, {\it ii)} a cylindrical accretion disk with radii as 
determined in PAPER~I (see also table~\ref{tab:elena}), and height 2H as already 
defined in subsection~\ref{sec:lflx}. The result is an accretion disk mass of 
M$_d\sim5.7\times10^{-15}M_\odot$, which is about 5 orders of magnitude smaller 
than 
the total mass accreted during the inter-outburst period 
($\sim10^{-9}M_\odot$, \ncite{Smak}). It is also $\sim$4 order of magnitude 
smaller 
than the critical mass (i.e. the maximum accretion disk mass before an 
outburst) expected in low mass transfer rate systems \footnote{We derived the 
value M$_{crit}\sim 4\times10^{-11} M_\odot$ either using average values for the 
accretion disk radii and surface density as predicted by  \scite{lasota} 
($R_{in}\sim4.5\times 10^9$cm, $R_{out}\sim 
1.1\times10^{10}$cm, $\log \Sigma 
\sim2.4$, for $\dot{M}\sim 10^{-11}M_\odot yr^{-1}$), or formula (18) in 
\scite{Smak} with $\alpha \sim 0.003$ as predicted for TOADs by \scite{HSC}.}. 
However, typical values for the mass transfer rates and 
critical masses have 
been computed using the assumption of $\alpha$-disk models which are optically 
thick and cannot therefore be compared directly with optically thin gas models. 
\scite{williams} predicts an accretion disk 
that is only partially optically 
thin at a mass transfer rate of $\dot{M} \sim 10^{-11}M_\odot yr^{-1}$, the 
optically thin region starting from a radius of $5\times 10^9$cm, outward. This 
value seems in quite good agreement with the radius of $r\sim2.6\times 10^9$cm 
at which we observe the transition of the Balmer line ratio from optically thick 
to optically thin case (see section 4 and figure 7 in PAPER~I). 

\begin{figure*}
\vspace{18cm}           
\caption{The radial velocity measurements and their best fit sine curves.  
Phasing is based on the photometric eclipse ephemeris of 
\protect\scite{skidthesis}, the period was fixed at the orbital period. while 
the amplitude, K1, and the phase $\phi_o$ were free parameters in the fit. The 
red-to-blue crossing of each line is shown as a vertical dashed line. The mean 
wavelength or systemic 
velocity, $\gamma$, is shown as a horizontal dotted 
line. }
\includegraphics{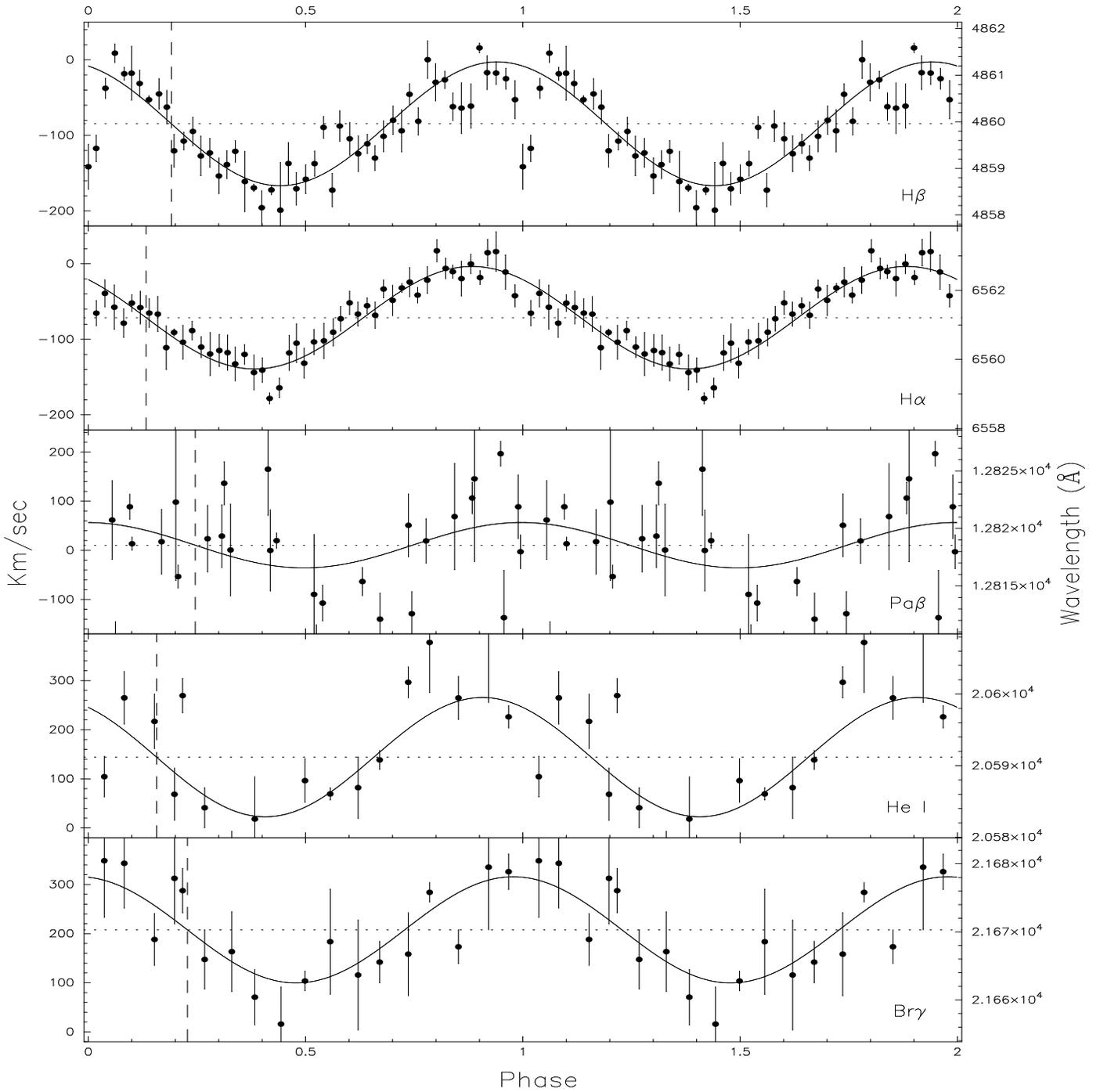}
\label{fig:rvplot}
\end{figure*}

\begin{table*}
\begin{minipage}{18cm}
\begin{center}
\caption{Results of fitting the radial velocity curves (See 
Fig~\ref{fig:rvplot}).}
\begin{tabular}{ccccc}
Line & Rest Wavelength & V/R crossing & $K_1$ amplitude & Systemic velocity \\
 & \AA & $\phi_o$ & km sec$^{-1}$ & ($\gamma$) km sec$^{-1}$ \\
\hline
H$\beta$ & 4861.33 & 0.1910$\pm$0.0058 & 82.0$\pm$2.5 & -84.6$\pm$2.0\\
H$\alpha$ & 6562.7725 & 0.1331$\pm$0.0061 & 67.9$\pm$2.8 & -71.4$\pm$1.9 \\
Pa$\beta$ & 12818.1 & 0.246$\pm$0.038 & 46.2$\pm$8.5 & 10.4$\pm$7.0 \\
He~I & 20581.3 & 0.157$\pm$0.014 & 121.4$\pm$12.6 & 144.1$\pm$9.2 \\
Br$\gamma$ & 21655.3 & 0.228$\pm$0.023 & 108$\pm$15 & 208$\pm$11 \\
\label{tab:rvfit}
\end{tabular}
\end{center}
\end{minipage}
\end{table*}

\section{Radial Velocity Curves}
\label{sec:radvel}

Optical and IR spectra were used to measure radial velocity curves for    
H$\beta$, H$\alpha$, Pa$\beta$, Br$\gamma$ and He~I ($\lambda2.06\mu$m) emission 
lines. Radial velocity curve measurements were not produced for the two emission 
lines He~I $\lambda1.08\mu$m and Pa$\gamma$ as they were strongly blended 
together. 
 
Figure~\ref{fig:rvplot} presents our radial velocity curves and their best fit 
for the five emission lines listed above. Each data point corresponds to the 
radial velocity of the emission line central wavelength. Central wavelengths 
were computed by using a technique similar to the Pogson's method to determine 
the mid-time of an eclipsing binary with an asymmetric eclipse \cite{pog}. The 
radial velocity measurements were fit by a grid-search method minimizing 
$\chi^2$. The sinusoidal fitting function used was:
\begin{equation}
v_i=\gamma+K_1\times \sin [2\pi(\phi_i+0.50-\phi_o)]
\label{eq:curvefit}
\end{equation}
where $v_i$ is the measured velocity and $\phi_i$ the observed photometric 
phase. The fitting parameters $\gamma$, K1, and $\phi_o$ are respectively  the 
systemic velocity, the Keplerian velocity of the white-dwarf around the system's 
center of mass, and the red-to-blue crossing in the radial velocity curve, 
corresponding to secondary inferior conjunction. We added the phase shift of 
0.50 in Eq.\ref{eq:curvefit} in order to have $\phi_o$ exactly matching the 
secondary star inferior conjunction, and not the primary star inferior 
conjunction. 

Our best fit parameters from each emission line represent an inconsistent 
set of values (see table~\ref{tab:rvfit}). The systemic velocity, $\gamma$, runs 
from negative to positive values, spanning a range of $\sim$ 300km sec$^{-1}$ 
wide. The white-dwarf Keplerian velocity, K1, runs from $\sim$46 to 120 km 
sec$^{-1}$; while the red-to-blue crossing, $\phi_o$, swings about phase 0.19, 
covering a range $\Delta\phi\simeq$0.1 in phase. Interestingly, a similarly 
inconsistent set of system parameters was determined for  the optical 
radial velocity curves of the SU~UMa type dwarf-nova VY~Aqr \cite{august}.  

When fitting and interpreting the sinusoidal radial velocity curves, it is 
assumed that: {\it i)} the disk orbits with Keplerian velocity around the 
white-dwarf, and {\it ii)} the accretion disk follows the white-dwarf motion 
around the binary's center of mass. These  approximations fail when hot-spot 
emission and disk asymmetries are not negligible. \scite{smak70}, shows that the 
hot-spot can induce a measured K1 larger than the true value and a spuriously 
delayed moment for inferior conjunction. \scite{ppt} explain a larger K1 
amplitude with non Keplerian motion.

We believe that our measured radial velocity curves result from combination of 
two different motions: {\it 1)} the motion of the accretion disk as described at 
points {\it i)} and {\it ii)} of the previous paragraph; and {\it 2)} the motion 
of the gas within the hot-spot region which moves along the ballistic trajectory 
at the stream velocity. Combination of the two velocity components in points 
{\it 1)} and {\it 2)} will change the amplitude and phase of a sinusoidal radial 
velocity curve, the largest departures occurring in systems where the hot-spot 
emission represents the largest fraction of the total emission line flux. 
An 
anisotropically emitting hot-spot may affect the $\gamma$ velocity 
measurements. Optically thin accretion disks are likely to be low density gas 
regions (see section~\ref{sec:corplot}), which neither affects the stream 
velocity of the in-falling gas, nor absorbs the flux emitted at the impact 
region. Thus, emission line profiles from optically thin accretion disks are 
expected to be heavily biased by the hot-spot emission.  

\begin{figure}
\vspace{8cm}
\includegraphics{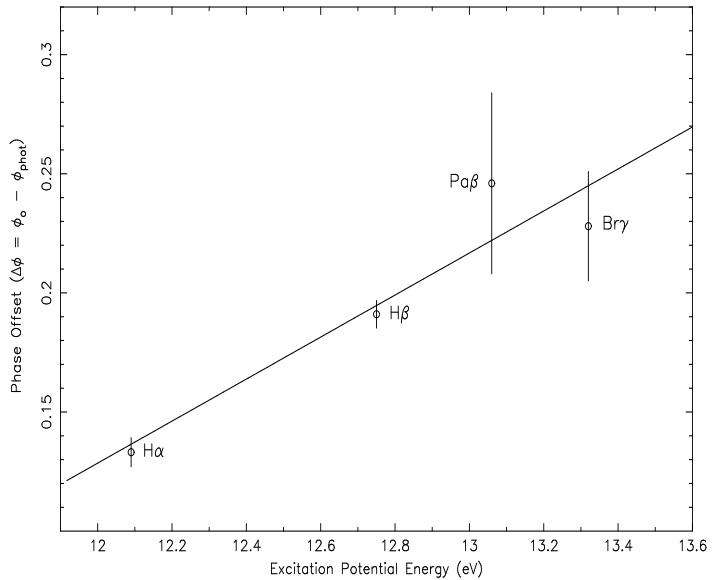}
\caption{The linear relationship between the observed phase offset and the 
excitation potential of the emission line. The straight line is just to 
highlight the linearity of the observed distribution. The slope is 
0.088(phase/eV); the y-intercept is -0.93(eV).}
\label{fig:david}
\end{figure}

\begin{table*}
\begin{center}
\small
\begin{tabular}{lccccc}
Star Name & DN Type& Orbital Period & Phase Offset & {\it i} & Ref. \\
 & & (days) & $\Delta\phi = \phi_o -\phi_{phot}$ & (deg) &  \\
\hline
OY Car	& SU & .06312 & .1167$\pm$ .0058 & 83.3$\pm$0.2$^\dagger$ & (1) \\
HT Cas & SU & .073647 &	.095$\pm$ .009 & 76.4 & (2) \\
Z  Cha	& SU &	.0745 &	.0904 $\pm$	.0017 & 81.7& (3) \\
EM Cyg	& ZC &	.29090950 & .057 $\pm$ .011 & 63$\pm$10$^\dagger$ & (4) \\
U  Gem	& UG &	.17708	& .015 $\pm$ .011 & 69.7$\pm$0.7$^\dagger$ & (5) \\
V2051 Oph  & UG?& .062427887 & .132 $\pm$ .020 & 80.5$\pm$2$^\dagger$ &	(6) \\
CN Ori	& UG &	.163190	& .03 $\pm$ .01 & 65$\rightarrow$70  & (7) \\
IP Peg	& UG & 	.15820	& .067 $\pm$ .008 & 79.3$\pm$0.9 & (8) \\
WZ~Sge	& SU &	.05668684707  & .1620 $\pm$ .0042 & 75$\pm$2 & (9) \\
DV UMa	& SU? &	.0859722  & .10 $\pm$ .02 & 72 & (10) \\
HS 1804+6753  &	UG & .2099370  & .044 $\pm$ .002 & 84.2$\pm$0.6 & (11) \\
\end{tabular}
\end{center}
\caption{\small Orbital period, phase offset, and inclination for the sample of 
dwarf-novae, see figure~\ref{fig:phoff}. References: (1) 
\protect\scite{schoembs83} $\&$ 
\protect\scite{bailey81}; (2) \protect\scite{young81}; (3) 
\protect\scite{marsh87b}; (4) \protect\scite{stoveretal}; (5) 
\protect\scite{smak76}; (6) \protect\scite{watts}; (7) 
\protect\scite{barrera89}; (8) \protect\scite{marsh88}; (9) this paper; (10) 
\protect\scite{szkody93}; (11) \protect\scite{fiedler97}. $^\dagger$ 
\protect\scite{kolb}.}
\label{tab:phoffsample}
\end{table*}

The phase offset is found to be correlated with the energy required to produce 
each emission line.
Figure~\ref{fig:david} shows the observed phase offset 
versus the excitation potential of each H emission line as given by the standard 
formula.  
Figure~\ref{fig:david} clearly shows a linear relationship between the observed 
phase offset and the emission line energy and implies a smeared out hot-spot 
region with decreasing excitation energy and temperatures as the gas moves 
down-stream. Results similar to those in figure~\ref{fig:david} may be obtained 
using the 
data in \scite{august} on VY~Aqr. We fit the phase offset computed by 
Augusteijn for three hydrogen emission lines and found a linear relationship 
with almost identical slope and y-intercept as in figure~\ref{fig:david}. 

The He~I 2.06$\mu$m emission line was not considered in figure~\ref{fig:david}. 
The He~I line 2.06$\mu$m is a transition strongly coupled with the resonant line 
at 584\AA. Both lines decay from the same 2$^1$P level but the 584\AA \ 
transition has a probability of occurrence which is $\sim10^3$ times larger 
\cite{naja}. Strong He~I emission at 2.06$\mu$m is rarely seen in astronomical 
objects and observed only in the extended atmosphere, wind and/or disk-like
formations of WN and Be super-giant stars, where they form because of the high 
temperature, velocity and density of the out-flowing gas (15000-30000K, 
$\sim$700 Km sec$^{-1}$, and  $\sim 10^{-5}$ M$_\odot$ yr$^{-1}$, respectively, 
\ncite{naja}). The presence of the He~I 2.06$\mu$m emission requires the He~I 
584\AA \ line to form in an optically thick, high temperature 15000-30000K 
region, which, in short period dwarf-novae, can only occur in the shock heated 
stream-disk impact region. Formation of He~I 2.06$\mu$m in a hot corona is 
excluded by the Doppler tomogram in PAPER~I figure 3.   

We interpret our results above to provide a picture for the quiescent accretion 
disk in WZ~Sge (figure~\ref{fig:geo}), where each hot-spot emission line is 
likely to be aligned along the stream trajectory into the accretion disk, with 
the higher energy emissions arising at the higher temperature outer edge of the 
accretion disk impact region, and the lower energy lines primarily forming 
further down-stream.

\subsection{Phase offset as a measure of the hot-spot bias}
\label{sec:corplot}

White-dwarf and/or hot-spot eclipses are present in the broad band photometric 
light curves of many dwarf-novae. In the case of the white-dwarf eclipse, the 
observed eclipse minimum defines the time of the secondary star inferior 
conjunction. In systems displaying only a bright spot eclipse, the phase offset 
between binary phase zero and the hot-spot eclipse is small ($\leq$0.05 
in phase, \ncite{greco}), so that we still have a good idea of the time of the 
true binary phase zero. Similarly, when measuring radial velocity curves,
we 
expect the spectroscopic ephemeris to match the photometric one in systems with 
perfectly symmetric accretion disk, and a phase offset larger than zero in 
systems having asymmetries in their disk. In particular, a larger phase offset 
is expected in systems where the hot-spot represents the main asymmetry in the 
accretion disk and a considerable fraction of the total emission line flux. This 
is the case of WZ~Sge given that our observations average out possible short 
term variation and Doppler tomograms in PAPER~I show accretion disk asymmetries 
always considerably fainter than the hot-spot emission; moreover in section 
\ref{sec:lp} we showed the accretion disk to be optically thin. To test the 
hypothesis of phase offset as a measure of the hot-spot bias and the optical 
properties of the accretion disk gas we searched the literature for measured 
dwarf-novae phase offsets and plot these values against the binary orbital 
period. Theory  predicts low mass transfer rates in short orbital period systems 
\cite{toad}, implying low density possibly optically thin accretion disks.

\begin{figure}
\vspace{8cm}
\includegraphics{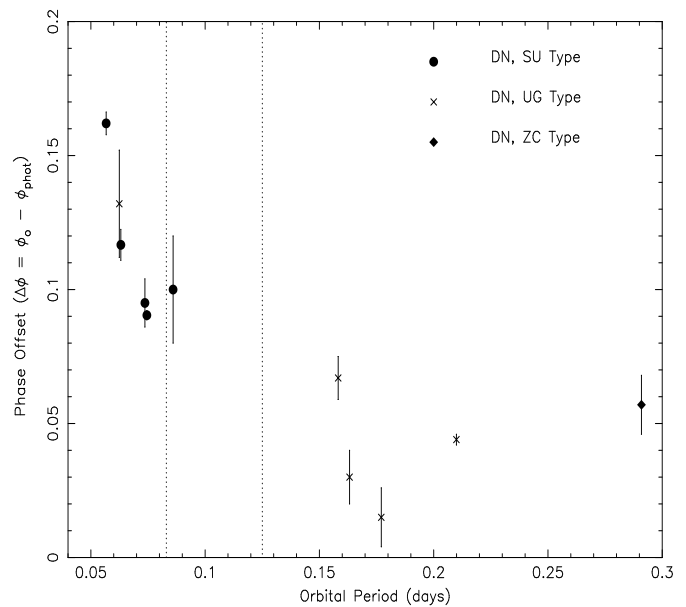}
\caption{The observed phase offset vs. orbital period. Phase offset is 
the delay of the spectroscopic secondary conjunction (red to blue crossing) as 
measured from the disk emission lines, with respect to the photometric eclipse 
minimum (i.e. the photometric phase zero). Different dwarf-novae subtypes are 
plotted with different symbols. Dotted vertical lines delimit the period-gap.}
\label{fig:phoff}
\end{figure}

Phase offsets have been observed in several different CVs (\ncite{watts}, 
\ncite{shafter92}, and \ncite{stover81}), but until now there has been no 
explicit or systematic search for its cause. \scite{watts}, observed phase 
offsets in both nova-like and SU UMa stars, claiming that the hot-spot cannot be 
the cause of such an offset. \scite{stover81} concluded that the best candidate 
to explain the phase offset observed in five dwarf-novae is only the hot-spot. 
Finally, \scite{shafter92} collected most of the observed phase offsets in the 
literature and plotted them vs. the binary orbital period but did not provide a 
possible explanation. 
 
The eleven dwarf-novae listed in table~\ref{tab:phoffsample} have been carefully 
selected in order to provide a homogeneous sample. This was achieved by 
selecting: {\it i)} high inclination dwarf-novae that show photometric eclipses, 
{\it ii)} only Balmer line radial velocity curves, and {\it iii)} publications 
reporting both photometric and spectroscopic analysis. Restrictions for point 
{\it i)} are to avoid biases other than the hot-spot and to reduce unknown 
inclination effects. Point {\it ii)} reduces the scatter in the measured phase 
offset by focusing on emission lines arising from regions with similar physical 
conditions, and finally, point {\it iii)} guarantees consistently measured phase 
offset and reduces any bias due to possible changes in the quiescent accretion 
state. Figure~\ref{fig:phoff} plots our selected sample which provides evidence 
for the expected correlation: the phase offset is larger in short orbital period 
systems. 

\section{Accretion disk size}
\label{sec:diskrad}

It is possible to estimate the inner and outer accretion disk radii in a 
dwarf-nova by measuring the emission line wing and peak separations 
respectively, and assuming a Keplerian velocity. For an annulus of gas in 
Keplerian motion at a distance $r$ from a white-dwarf of mass M$_{WD}$, orbiting 
at an angle $i$ to the line of sight, the Keplerian velocity is:
\begin{equation}
V_{kep}=\sqrt{{GM_{WD} \over r}} \sin i
\label{eq:vkep}
\end{equation}

In a double peaked emission line from an accretion disk, the Keplerian velocity 
of the outer disk radius corresponds to half the line peak separation (hereafter 
HPS); while, the Keplerian velocity at the inner accretion disk radius 
corresponds to half the line wing separation (hereafter HWS).

It is important both to have reasonable system parameters and to  measure HWS 
and HPS on unbiased spectra, i.e. spectra in which the wings and peaks of the 
accretion disk emission line are not affected by hot-spot emission 
(figure~\ref{fig:specplot}, phase 0.3 and 0.84). Spectra in which the hot-spot 
emission clearly contaminates wings and/or peaks of the accretion disk emission 
line profile  may provide a measure of the hot-spot bias. Both HWS and HPS were 
measured reading the cursor position with the IRAF task {\it splot}. HWS values 
from each spectra are the average over 5 measurements. 

\begin{table*}
\begin{minipage}{18cm}
\begin{center}
\begin{tabular}{cccccc}
Line & \multicolumn{2}{c}{Uncontaminated Spectra} & \multicolumn{2}{c}{Average 
Spectra} & Ref. \\
\hline
& HPS &  HWS & HPS &  HWS  & \\
& (km sec$^{-1}$) & (km sec$^{-1}$)  & (km sec$^{-1}$) & (km sec$^{-1}$) & \\
\hline
H$\alpha$ & $711 \pm 10$ & $1606 \pm 65$ & $676 \pm 19$ & $2133 \pm  6$ & (1) \\
H$\beta$  & $762 \pm 28$ & $1692 \pm 98$  & $647 \pm 15$ & $1745 \pm  23$ & (1) 
\\
Br$\gamma$ & $702 \pm 66$ & $1598 \pm 50$ & $695 \pm 14$ & $1709 \pm 13$ & (1)\\
He~I 2.06$\mu$m & $608 \pm 66$ & $1740 \pm 97$ & $624 \pm 15$ & $2120 \pm 74$ 
& (1) \\ 
P$\beta$ & $619 \pm 154$ & $1868 \pm 260$ & $657 \pm 23$ & $2109 \pm 74$ & (1)\\ 
Pa$\gamma$ & & & $633 \pm 27$ & -- & (1) \\
He~I 1.08$\mu$m & & & $646 \pm 28$ & -- & (1) \\
H$\alpha$ & & & $731 \pm 180$ & $\sim2300 \pm 180$ & (2) \\
\end{tabular}
\end{center}
\caption{Half peak separation (HPS) and half wing separation (HWS) as measured 
in both uncontaminated spectra and averaged spectra. Selection criteria for 
uncontaminated spectra are described in the text. IR measures are less certain 
due to low S/N data and have not been considered in accretion disk radii 
computations (see text). References: (1) this paper, (2) 
\protect\ncite{Gill}.}
\label{tab:hpews}
\end{minipage}
\end{table*}

\begin{table*}
\begin{minipage}{18cm}
\begin{center}
\scriptsize
\begin{tabular}{llccccccccccc}
$M_{wd}$ & $R_{wd}$ & q & {\it i $^o$} & \multicolumn{3}{c|}{R$_{out}$ 
($\times10^{10}$cm)} & \multicolumn{3}{|c|}{R$_{in}$ ($\times10^{10}$cm)} & 
\multicolumn{2}{|c|}{Fraction R$_{L1}$} & Ref. \\
($M_{\odot}$) & $\times10^{10}$cm & $\frac{M_2}{M_1}$ & & H$\alpha$ & H$\beta$ 
& Br$\gamma$ & H$\alpha$ & H$\beta$ & Br$\gamma$ & H$\alpha \& Br\gamma$ & 
H$\beta$ & \\ 
\hline
$0.45\pm 0.19$ & $0.098\pm 0.021$ & 0.13 & 75 & $1.10\pm0.03$ & $0.96\pm 0.07$ & 
1.13$\pm$0.21 & $0.22\pm 0.02$ & $0.19\pm 0.02$ & $0.22 \pm 0.01$ & 0.53 & 0.48 
& (1) \\
$1.2\pm 0.25$ & 0.039 & 0.05 & 77 & $2.99\pm 0.09$ & $2.60\pm 0.19$ & 
3.07$\pm$0.58 & $0.59\pm 0.05$ & $0.53\pm0.06$ & $0.59 \pm 0.04$ & 0.93 & 0.84 & 
(2) \\
$0.3$ & 0.055 & 0.19 & 75 & $0.73\pm 0.02$ & $0.64\pm 0.05$ & 0.75$\pm$0.14 & 
$0.14\pm 0.01$ & $0.13\pm 0.01$ & $0.15\pm 0.01$ & 0.43 & 0.38  & (3) \\
$1.1$ & ~~- & 0.05 & 80 & $2.78\pm0.08$ & $2.43\pm0.18$ & 2.87$\pm$0.54 &  
$0.55\pm0.04$ & $0.49\pm0.06$ & $0.55\pm 0.03$ & 0.90 & 0.81  & (4) \\
0.82$\pm$0.10 & ~-~ & 0.07 & 75 & 1.94$\pm$0.06 & 2.04$\pm$0.08  & - & 
0.42$\pm$0.26 & 0.26$\pm$0.17 & - & 0.63  &  & (5) \\
\end{tabular}
\caption{The disk parameters determined from the line peak and line wing 
separations of the uncontaminated spectra. Values for $q$ assume  
$M_2=0.058M_\odot$ 
\protect\cite{Smak}. Columns 11 and 12 are the outer disk radii normalized to 
the 
primary Roche lobe radius. References: (1) Smak (1993), (2) Spruit \& 
Rutten (1998), (3) Cheng et al. (1997), (4) Gilliland, Kemper and Suntzeff 
(1986), (5) PAPER~I. Values from PAPER~I were computed applying a different 
method, i.e. assuming that the outer radius extends to the 3:1 tidal resonance 
radius and deriving M$_{WD}$, R$_{in}$, and R$_{out}$. The orbital inclination 
were taken from Smak (1993). Inner and Outer radii corresponding to the HWS and 
HPS measured on the He~I and Pa$\beta$ emission lines were not computed because 
of their larger uncertainties and the fact that they are statistically identical 
to the H$\alpha$ and Br$\gamma$ measurements. }
\end{center}
\label{tab:elena}
\end{minipage}
\end{table*}

\subsection{Measurements on uncontaminated spectra}
\label{sec:uncontaminated}

We selected a sample of uncontaminated phase binned H$\alpha$ and H$\beta$ 
spectra. The assumption is that the hot-spot emission does not affect the wings 
of the accretion disk emission lines. This selection criteria produced three 
binned H$\alpha$ spectra (at phases 0.28, 0.30 and 0.32) and six binned H$\beta$ 
spectra (at phases 0.28, 0.30, 0.32, 0.34, 0.82 and 0.84). We applied the same 
selection criteria to IR spectra and selected two uncontaminated spectra in both 
the J and K band (at phases 0.33 and $\sim$0.84 in each case). In case the 
hot-spot feature was not evident (figure~\ref{fig:irsample}, J-band, phase 0.3) 
we select those spectra which are the closest in phase to the uncontaminated 
optical spectra listed above. HWS and HPS were first measured on each bin 
separately, and then averaged. HWS and HPS measured values are in table 
\ref{tab:hpews}. We see that values from optical and IR emission lines are in 
good agreement each other, the exceptions being the He~I $2.06 \mu$m and 
Pa$\beta$ lines which apparently have smaller  HPS value and larger HWS values 
than in the H$\alpha$, H$\beta$, and Br$\gamma$ measurements. We applied a 
t-student test (to a confidence level of $\alpha = 0.05$), to check for 
statistical differences between pairs of values. The test showed that the HWS 
values have to be considered statistically identical in all the emission lines, 
i.e. the same inner disk radii; while the H$\beta$ HPS is statistically larger 
than the measured HPS from H$\alpha$, Br$\gamma$, He~I $2.06 \mu$m, and 
Pa$\beta$, thus implying a smaller outer disk radius ($\sim 10 \% $).

Velocities from uncontaminated spectra were used in equation \ref{eq:vkep} to 
compute the accretion disk radii for a variety of primary masses and 
inclinations (see table~\ref{tab:elena}). It is evident that small white-dwarf 
masses (M$_{WD}<0.5$M$_\odot$), imply a small accretion disk outer radius 
($<50\%$ of the primary Roche lobe radius) and a relatively large ratio 
R$_{WD}$/R$_{in}$ ($<0.5$); while larger masses (M$_{WD}\geq 1$M$_\odot$) give a 
large outer disk radius (up to 90\% of the primary Roche lobe radius) and 
R$_{WD}$/R$_{in}\sim0.07$, implying an annulus shaped accretion disk. The input 
value of the white-dwarf mass does not affect the ratio, R=R$_{in}$/R$_{out}$, 
which is equal to 0.2 in each case. 

\begin{table}
\begin{center}
\begin{tabular}{ccc}
\multicolumn{3}{c}{Half Wing Separation (Km sec$^{-1}$)}\\
\hline
& Uncontaminated Spectra$^\dagger$ & Averaged Spectra \\
  Line    &  & K1 subtracted \\
\hline
H$\alpha$ &  $1606 \pm 65$ & $2065 \pm 7$\\
H$\beta$ &  $1692 \pm 98$ & $1663 \pm 23$\\
Pa$\gamma$ & $1868 \pm 260$ & $2063 \pm 75$ \\
He~I (2.06$\mu$m) & $1740 \pm 97$ & $1999 \pm 75$\\
Br$\gamma$ & $1598 \pm 50$ & $1601 \pm 20$\\
\end{tabular}
\end{center}
\caption{Half wing separation value corrected for the white-dwarf orbital 
motion. Each emission line measure has been corrected with the K1 velocity 
determined in the corresponding radial velocity curve. $^\dagger$ Same values as 
 given in table~\ref{tab:hpews}. }
\label{tab:k1corr}
\end{table}

\subsection{Measurements on the averaged spectra }
\label{sec:total}

Optical and IR average spectra show relatively symmetric accretion disk emission 
lines. This is expected because the hot-spot motion, i.e. the S-wave, smoothes 
out the hot-spot features at each phase, and evenly distributes its emission 
throughout the orbit. Average spectra also show broader emission lines than 
single spectra. This happens for two reasons: {\it i)} the common orbital motion 
of the accretion disk and the white-dwarf around the binary center of mass will 
broaden the disk emission line by a term K1, i.e. the white-dwarf Keplerian 
velocity; and {\it ii)} the wings of the hot-spot emission blend with the blue 
and red-shifted accretion disk emission lines. The S-wave velocity is usually 
smaller than the outer disk edge Keplerian velocity (\ncite{Gill}; hereafter 
GKS), thus, we expect average spectra with broader accretion disk emission peaks 
and smaller peak separation than observed in single spectra. HPS and HWS from 
average spectra are given in table~\ref{tab:hpews}. HPS values from optical and 
IR emission lines agree fairly well and confirm  --with the exception of He~I 
$2.06 \mu$m and Pa$\beta$-- the expectation. HWS values are in good agreement 
too, emission lines with larger HWS value are probably affected by high velocity 
components from the hot-spot. HWS and HPS measured in the averaged spectra can 
give information about the S-wave bias when compared with wing and peak 
separations from uncontaminated spectra. They also test the assumptions used in 
the accretion disk radii computations. 

We compared HWS values from averaged and uncontaminated spectra after correction 
for the white-dwarf orbital motion. HWS from uncontaminated spectra, and HWS 
from averaged spectra, K1 subtracted, should be equal in the absence of hot-spot 
high velocity components. Table~\ref{tab:k1corr} shows that HWS values from 
uncontaminated spectra and K1 subtracted averaged spectra agree very well each 
other in the case of H$\beta$ and Br$\gamma$ emission lines; while H$\alpha$ and 
He~I $2.06 \mu$m have HWS values from K1 subtracted average spectra which are 
$\sim$ 450 and 250 Km sec$^{-1}$ larger than expected in the absence of high 
velocity components in the hot-spot. Such components bias both radial velocity 
curve measures and the accretion disk radii computation. The fact that we 
observe the largest hot-spot high velocity component in the H$\alpha$ line is 
consistent with figure~\ref{fig:geo} and the less energetic emission  forming 
downstream in the accretion disk. High velocity hot-spot components in the 
Pa$\beta$ emission line should be taken cautiously, because of larger 
uncertainties in the measurement. 

It is worth noting that each average spectra was K1 subtracted by the 
correspondent white-dwarf orbital motion as determined in section 
\ref{sec:radvel}. The K1 values from radial velocity curve fitting are biased by 
the hot-spot and do not reflect the true orbital motion of the white-dwarf. 
However, because the hot-spot bias is believed to increase the true K1 value 
(see section~\ref{sec:radvel}, \ncite{smak70}, and \ncite{ppt}), then the 
applied correction to our averaged spectra should be over-estimated. 

Table~\ref{tab:hpews} also reports HWS and HPS as we measured by hand from the 
average 
H$\alpha$ spectra in figure 1 of GKS. Comparison of the GKS 
determinations with those from our averaged spectra gives a chance to check for 
changes in the quiescent accretion state of WZ~Sge. The data in GKS 
were 
obtained almost one year after the last outburst of WZ~Sge in Dec 1978 
and ours 
were obtained about 18 years after the outburst. We measured larger HWS and HPS 
values in the GKS spectra than in our averaged spectra. In particular, in the 
GKS spectra the HWS is $\sim$200 Km~sec$^{-1}$ larger than in our averaged 
spectra; while, the HPS matches the values we measured on our uncontaminated 
spectra. Therefore, we may conclude that: {\it i)} the H$\alpha$ hot-spot 
emission high velocity component was possibly larger one year after the 
super-outburst than 18 years later; {\it ii)}  the hot-spot emission was 
probably weaker at the time of GKS observations such that it didn't affect 
the accretion disk peak separation; {\it iii)} the peak separation in GKS 
spectra may be considered free from hot-spot bias and implies constant outer 
accretion disk radius in the H$\alpha$ emission. 

Particularly interesting is the conclusion of point {\it iii)} which claims a 
constant accretion disk outer radius. Indeed, the same HPS value has been 
observed in all quiescent spectra of WZ~Sge obtained in the last 40 years 
(see table~\ref{tab:voutersold}). One possible explanation is in the assumption 
of the outer accretion disk radius at the 3:1 tidal resonance radius. A constant 
accretion disk outer radius has some important implications relating to current 
outburst theory. Some disk outburst theories \cite{Meyer98,wynn99} predict that, 
during quiescence, the outer radius slowly increases up to the 3:1 resonance 
radius. Once reached  at the 3:1 resonance, quiescent super-humps should be 
present. 

\begin{table}
\caption{Results of measurements of the outer disk velocity spanning 40 years.  
Outbursts occurred in June 1946 and Dec 1978. $^\dagger$ years since the 
previous outburst. References: (1) \protect\scite{greenstein}, (2) 
\protect\scite{KK}, (3) \protect\scite{Gill}, (4) \protect\scite{Smak}, (5) 
\protect\scite{neustroev}, (6) This paper.}
\begin{center}
\begin{tabular}{clrlc}
Line & Date & $\Delta$yr$^\dagger$ & $v_{outer}$ (km/sec) & Ref.\\
\hline
H$\beta$ & 1956 Aug 15 & 10 & 720 & (1) \\
H$\gamma$ & 1956 Aug 15 & 10 & 710 & (1) \\
H$\beta$ & 1962 Aug 4 & 16 & 720$\pm$14 & (2) \\
H$\gamma$ & 1962 Aug 4 & 16 & 720$\pm$14 & (2) \\
H$\alpha$ & 1979 July 16 & 1 & 731 & (3) \\
H$\beta$ & 1983 Nov 24-28 & 5 & 730$\pm$30 & (4) \\
H$\gamma$ & 1983 Nov 24-28 & 5 & 730$\pm$30 & (4) \\
H$\alpha$ & 1994 May 30 & 16 & 747$\pm$40 & (5) \\
H$\alpha$ & 1996 July 27/28 & 18 & 711$\pm$10 & (6) \\
H$\beta$ & 1996 July 27/28 & 18 & 762$\pm$28 & (6) \\
\end{tabular}
\end{center}
\label{tab:voutersold}
\end{table}

\begin{table*}
\begin{minipage}{18cm}
\begin{center}
\caption{Our measurements of the disk parameters from the uncontaminated 
spectra using the method described in Smak (1981) compared to the measurements 
of Mennickent \& Arenas (1998). Measurements made assuming a Keplerian 
disk $\dagger$, and the data of Neustroev (1998). References: (1) 
\protect\scite{Menn}, (2) This paper, (3) \protect\scite{neustroev}.}
\label{tab:lastt}
\begin{tabular}{ccccccc}
 Line & $A_{84}$ & $A_{41}$ & U & $\alpha$ & R & Ref.\\ \hline
 H$\alpha$ & 0.07$\pm$0.02 & 0.104$\pm$0.001 & 0.12 & 0.5 & 0.30 & (1) \\
 H$\alpha$ & 0.072 & 0.113 & 0.013 & $>$1 & $>$0.2 & (2) \\
 H$\beta$ & 0.077 & 0.122 & 0.016 & $>$0.75 & $>$0.23 & (2) \\
 H$\alpha$ & & & & & 0.20 & (2)$\dagger$ \\
 H$\beta$ & & & & & 0.20 & (2)$\dagger$ \\
 H$\alpha$ & & & & 1.66$\pm$0.19 & 0.08$\pm$0.04 & (3) \\
\end{tabular}
\end{center}
\end{minipage}
\end{table*}
 
\subsection{Smak's method of determining R$_{in}$/R$_{out}$ }
\label{sec:smakmeth}

The accretion disk radii ratio, R=R$_{in}$/R$_{out}$, is often the only value 
provided by authors because of the large uncertainties affecting the 
determination of the orbital inclination and the binary star masses. A commonly 
applied method to compute R is the one developed by \scite{Smak81}.

Smak's method assumes axially symmetric accretion disk in Keplerian motion and a 
power law flux distribution ($f\sim r^{-\alpha}$) for the disk emission. The R 
and $\alpha$ values depend on the parameters U, A$_{84}$, and A$_{41}$ which are 
defined as follow: U=$res/\Delta\lambda$, with $res$=instrumental resolution, 
A$_{84}\equiv\log W_{0.8}-\log W_{0.4}$, and A$_{41}\equiv\log W_{0.4}-\log 
W_{0.1}$, where W$_{0.8}$, W$_{0.4}$, and W$_{0.1}$ are the emission line width 
at the fractions 0.8, 0.4, and 0.1 of the peak height above the continuum, 
respectively. 

\scite{Menn} applied Smak's method to WZ~Sge and determined A$_{84}=0.07$, 
A$_{41}=0.1$, and R=0.3 (see table~\ref{tab:lastt}). A ratio R=0.3 is 50\% 
larger than the value 0.2 we determined in section~\ref{sec:uncontaminated}. We 
investigated such a difference to understand whether it reflects changes in the 
quiescent accretion disk size or it implies inconsistencies between the Smak's 
method and the method described in section~\ref{sec:uncontaminated}. We applied 
Smak's method to our H$\alpha$ and H$\beta$ uncontaminated spectra and found 
A$_{84}$ and A$_{41}$ values matching those by Mennickent and Arenas (see 
table~\ref{tab:lastt}). However, we expect a ratio $0.2 \leq R < 0.3$. Then, 
we 
may conclude that: {\it i)} the two methods are consistent and, in 
particular, the assumptions of axisymmetric disk and power law flux distribution 
required by the Smak's method are not fundamental to the determination of R and 
Smak's 
method may be applied also to  asymmetric accretion disks such as WZ 
Sge; {\it ii)} the two accretion disk radii ratio, R=0.3 by Mennickent and 
Arenas and R=0.2 by us, may correspond to a real change in WZ~Sge's quiescent 
accretion state and indicate an inward motion of the inner accretion disk radius 
toward the white-dwarf surface (in section~\ref{sec:total} we showed that the 
outer disk radius remains constant during quiescence).  

\section{Summary and Conclusions}
\label{sec:conclusion}

We observed both hot-spot and accretion disk line emission to vary in shape and 
strength throughout an orbital period, and conclude that both the hot-spot and 
the accretion disk are asymmetric and anisotropically emitting. We determined 
the gas at the impact region to have just the stream velocity, and the accretion 
disk is a low density gas with little drag effects on the in-falling material 
from the secondary star. There are different Balmer decrements at the impact 
region and in the rest of the accretion disk, and we conclude that the gas has a 
different opacity, $\tau$, within the two regions. In particular, the hot-spot 
is optically thick in the lines, while the accretion disk appears to be   
optically thin in the lines. 

Our measured radial velocity curves, from five emission lines in the optical and 
in the infrared are found to present an inconsistent set of system parameters. 
We explain this result as a bias due to the hot-spot emission. In particular, we 
showed  that the hot-spot emission delays the apparent time of the secondary 
inferior conjunction with respect to photometric phase zero. We also show that 
the phase offset depends both on the excitation potential energy of the 
considered emission line and on the optical depth of the gas in the accretion 
disk. Our extended analysis of a selected sample of dwarf-novae shows increasing 
phase offset with shorter orbital period, consistent with the idea of optically 
thin accretion disks in low mass transfer rate systems.

Summarizing our results from the emission line profile analysis and radial 
velocity curve computations presented here, we can formulate a probable 
structure for the WZ~Sge accretion disk. The accretion disk is generally a low 
density, optically thin gas. The accretion disk is neither symmetric, nor has it 
a uniform gas density or temperature structure (see also PAPER~I). The hot-spot 
is 
optically thick in the emission lines, not symmetric in shape, and not an 
isotropic emitter. The hot-spot emission arises from a multicomponent extended 
region in which each emission component is not visible at all phases and the 
hot-spot emission varies in strength throughout the orbital period. These 
variable hot-spot emissions lead to an extended hot-spot region which does not 
emit strongly when viewed from the down-stream direction (phase 0.5 and 
following), and shows a temperature gradient along the stream trajectory.  
Figure~\ref{fig:geo} summarizes our conclusion. 

The qualitative picture of WZ~Sge provided by our observations and data analysis 
shows a peculiar quiescent accretion disk dissimilar to others observed at 
present. WZ~Sge is the first short orbital period dwarf-nova showing evidence
of optically thin emission both in the continuum (\ncite{Ciardi98}, and 
\ncite{skidthesis}), and the lines. Multicolor photometry of OY~Car, Z~Cha, and 
HT~Cas (\ncite{janet92}, \ncite{janet90}, \ncite{marsh87a}) shows evidence of 
quiescent accretion disk optically thin in the continuum but thick in the 
emission lines. This work establishes a first step toward an understanding of 
the spatial flux and material distribution within the WZ~Sge accretion disk and 
other TOAD (Tremendous Outburst Amplitude Dwarf-novae, \ncite{toad}) candidates.

The accretion disk radii determined in sec.\ref{sec:uncontaminated} show that 
determination of the inner and outer radius strongly depend on the assumed 
white-dwarf mass and the binary orbital inclination. Unfortunately, white-dwarf 
mass and radius determinations are still affected by large uncertainties. The 
uncertainties on the WZ~Sge white-dwarf parameters prevent us from uniquely 
determing the actual size of the accretion disk, the fraction of the primary 
Roche lobe it fills, or its exact shape. Whether the accretion disk extends down 
to the white-dwarf surface or is a ring-like accretion disk as suggested by some 
previous studies (\ncite{Meyer98}, \ncite{Menn}, \ncite{steve}), cannot be 
uniquely determined. We determined the accretion disk radii ratio in WZ~Sge, 
R=R$_{in}$/R$_{out}$, to be 0.2, and observed that the outer accretion disk 
radius does not vary during the quiescence period.  One explanation for the 
constant accretion disk outer radius is the assumption that the accretion disk 
extends to the 3:1 tidal resonance radius. One possible test of this idea would 
be to search for and find quiescent super-humps. 

\section*{Acknowledgments}
The optical data used in this study were obtained by Henke Spruit and Rene 
Rutten with the William Herschel Telescope of the ING at La Palma. The William 
Herschel Telescope is operated on the island of La Palma by the Isaac Newton 
Group in the Spanish Observatorio del Roque de los Muchachos of the Instituto 
the Astrophysica de Canarias. The United Kingdom Infrared Telescope is operated 
by the Joint Astronomy Centre on behalf of the U.K. Particle Physics and 
Astronomy Research Council. SBH acknowledges partial support of this research 
from the NSF grant AST 98-19770 and from the University of Wyoming office of 
research. We thank Dr. Peter Tamblyn for his help and information concerning 
He~I emission.


\begin{thebibliography}{{Skidmore, Mason, Howell, Littlefair, Dhillon \& 
Ciardi}{1999}}

\bibitem[\protect\citename{Augusteijn }1994]{august}Augusteijn~T., 1994, A\&A, 
292, 481

\bibitem[\protect\citename{Bailey \& Ward }1981]{bailey81}Bailey~M.E., 
Ward~M., 
1981, MNRAS, 196, 425

\bibitem[\protect\citename{Barrera \& Vogt }1989]{barrera89}Barrera~L.H., 
Vogt~N., 1989, A\&A, 220, 99

\bibitem[\protect\citename{Cheng {\rm et~al. } }1997]{cheng}Cheng~F.H., 
Sion~E.M., Szkody~P., Huang~M., ApJ, 484, 149

\bibitem[\protect\citename{Ciardi {\rm et~al. }}1998]{Ciardi98}Ciardi~D.R., 
Howell~S.B., Haushchildt~P.H., Allard~F., 1998, ApJ, 504, 450

\bibitem[\protect\citename{Fiedler, Barwig \& Mantel 
}1997]{fiedler97}Fiedler~H., Barwig~H., Mantel~K.H., 1997, A\&A, 327, 173

\bibitem[\protect\citename{Drake \& Ulrich }1980]{darke}Drake~S.A., Ulrich~R.K., 
1980, ApJSS, 
42, 351

\bibitem[\protect\citename{Gilliland, Kemper \& Suntzeff 
}1986]{Gill}Gilliland~R., Kemper~E., Suntzeff~N., 1986, (GKS), ApJ, 301, 252

\bibitem[\protect\citename{Glasby }1970]{pog}Glasby~I.S., 1970, {\it The Dwarf 
Novae}, American Elesvier Publishing Company Inc., New York

\bibitem[\protect\citename{Greenstein }1957]{greenstein}Greenstein~J., 1957, 
ApJ, 126, 23

\bibitem[\protect\citename{Hantzios }1988]{greco}Hantzios~P.A., 1988, PhD 
thesis, Ohio State University 

\bibitem[\protect\citename{Horne \& Marsh }1986]{Horne86b}Horne~K., Marsh~T.R., 
1986, MNRAS, 218, 761

\bibitem[\protect\citename{Howell, Rappaport \& 
Politano }1997]{toad}Howell~S.B.,Rappaport~S., Politano~M., 1997, MNRAS, 287, 
929 

\bibitem[\protect\citename{Howell, Szkody \& 
Cannizzo }1995]{HSC}Howell~S.B.,Szkody~P., Cannizzo~J.K., 1995, ApJ, 439, 
337  

\bibitem[\protect\citename{Howell {\rm et~al. }}1999]{steve}Howell~S.B.,
Ciardi~D.R., Szkody~P., Van~Paradijs~J., Kuulkers~E., Cash~J., Sirk~M., 
Long~K., 1999, PASP, 111, 342

\bibitem[\protect\citename{Krzeminski \& Kraft }1964]{KK}Krzeminski~W., 
Kraft~R., 1964, ApJ, 140, 921

\bibitem[\protect\citename{Lasota, Hameury \& Hur$\acute{e}$ 
}1995]{lasota}Lasota~J.P., Hameury~J.M., Hur$\acute{e}$~J.M., 1995, ApJSS, 53, 
523

\bibitem[\protect\citename{Lubow \& Shu }1976]{lu76}Lubow~S.H., Shu~F.H., 1976, 
ApJLett, 207, L53

\bibitem[\protect\citename{Marsh }1987]{marsh87a}Marsh~T., 1987, MNRAS, 
228, 779


\bibitem[\protect\citename{Marsh, Horne \& Shipman }1987]{marsh87b}Marsh~T.R., 
Horne~K., Shipman~H.L., 1987, MNRAS, 225, 551

\bibitem[\protect\citename{Marsh }1988]{marsh88}Marsh~T., 1988, MNRAS, 
231, 1117

\bibitem[\protect\citename{Mennickent \& Arenas }1998]{Menn}Mennickent~R., 
Arenas~J., 1998, PASJ, 50, 333

\bibitem[\protect\citename{Meyer-Hofmeister, Meyer \& Liu 
}1998]{Meyer98}Meyer-Hofmeister~E., Meyer~F. \& Liu~B., 1998, A\&A, 339, 507

\bibitem[\protect\citename{Najarro {\rm et~al.} }1994]{naja}Najarro~F., 
Hillier~D.J., Kudritzki~R.P., Krabbe~A., Genzel~R., Lutz~D., Drapatz~S., 
Geballe~T.R., 1994, A\&A, 285, 573

\bibitem[\protect\citename{Neustroev }1998]{neustroev}Neustroev~V., 1998, 
ARep, 42, 748

\bibitem[\protect\citename{Paczynski, Piotrowski \& 
Turski }1968]{ppt}Paczynski~B.,Piotrowski~S.L.,Turski~W., 1968, ASS, 
2, 254

\bibitem[\protect\citename{Patterson {\rm et~al. }}1996]{patt96}Patterson~J., 
Augustejin~T., Harvey~D.A., Skillman~D.R., Abbott~T.M.C.,Thotstensen~J., 
1996, PASP, 108, 748

\bibitem[\protect\citename{Ritter \& Kolb }1998]{kolb}Ritter~H., Kolb~U., 
1998, A\&AS, 129, 83

\bibitem[\protect\citename{Skidmore }1998]{skidthesis}Skidmore~W., 1998, PhD 
thesis, University of Keele

\bibitem[\protect\citename{Skidmore {\rm et~al. }}2000]{pap1}Skidmore~W., 
Mason~E., Howell~S.B., Ciardi~D.R., Littlefair~S., Dhillon~V.S., 2000, MNRAS, 
accepted

\bibitem[\protect\citename{Shafter }1991]{shafter92}Shafter~A.W., 1991, {\it 
Fundamental Properties of Cataclysmic Variable Stars}, Proceeding of the 
12$^{th}$ North American Workshops on Cataclysmic Variables, 39

\bibitem[\protect\citename{Schoembs \& Hartmann }1983]{schoembs83}Schoembs~R., 
Hartmann~K., 1983, A\&A, 128, 37 

\bibitem[\protect\citename{Smak }1970]{smak70}Smak~J., 1970, Act Ast, 20, 311

\bibitem[\protect\citename{Smak }1976]{smak76}Smak~J., 1976, Act Ast, 26, 277

\bibitem[\protect\citename{Smak }1981]{Smak81}Smak~J., 1981, Act Ast, 31, 395

\bibitem[\protect\citename{Smak }1993]{Smak}Smak~J., 1993, Acta Ast, 43, 212

\bibitem[\protect\citename{Spruit \& Rutten }1998]{spruit1998}Spruit~H., 
Rutten~R., 1998, MNRAS, 299, 768

\bibitem[\protect\citename{Stover }1981]{stover81}Stover~R.J., 1981, ApJ, 
249, 673

\bibitem[\protect\citename{Stover, Robinson \& Nather 
}1981]{stoveretal}Stover~R.J., Robinson~E.L., Nather~R.E., 1981, ApJ, 248, 696

\bibitem[\protect\citename{Szkody \& Howell }1993]{szkody93}Szkody~P., 
Howell~S.B., 1993, ApJ, 403, 743

\bibitem[\protect\citename{Young, Schneider \& Shectman 
}1981]{young81}Young~P., Schneider~D.P., Shectman~S.A., 1981, ApJ, 245, 1043

\bibitem[\protect\citename{Warner }1995]{warner95}Warner~B., 1995, {\it 
Cataclysmic Variable Star}, 691 Cambridge University Press, Cambridge

\bibitem[\protect\citename{Watts {\rm et~al. }}1986]{watts}Watts~D., 
Bailey~J., 
Hill~P., Greenhill~J., McCowage~C., Carty~T., 1986, A\&A, 154, 197

\bibitem[\protect\citename{Williams }1980]{bob}Williams~R.E., 1980, ApJ, 
235, 939

\bibitem[\protect\citename{Williams }1983]{williams83}Williams~G.A., 1983, 
ApJSS, 53, 523

\bibitem[\protect\citename{Williams \& Shipman }1988]{will88}Williams~G.A., 
Shipman~H.L., 1988, ApJ, 326, 738

\bibitem[\protect\citename{Williams  }1991]{williams}Williams~G.A., 1991, AJ, 
101, 1929

\bibitem[\protect\citename{Wood,  }1990]{janet90}Wood~J.H., 1990, MNRAS, 243, 
219

\bibitem[\protect\citename{Wood, Horne \& Vennes  }1992]{janet92}Wood~J.H., 
Horne~K., Vennes~S., 1992, ApJ, 385, 294

\bibitem[\protect\citename{Wynn {\rm et~al. }}2000]{wynn99}Wynn~G.{\rm 
et~al.}, 2000, NewAR, in prep.

\end{thebibliography}
\end{document}